\documentstyle[prd,aps,12pt,amssymb,preprint]{revtex}
\begin{document}
\draft
\textheight=8.5 in
\textwidth=6.0 in
\tighten
\date{October, 1993}
\preprint{RUB-TPII-57/93 \\}
\title{NUCLEON DISTRIBUTION AMPLITUDES AND OBSERVABLES \\}
\author{ N. G. STEFANIS
\thanks{Invited talk presented at the International Workshop on Hadron
Structure '93, Bansk\'a \v Stiavnica, Slovakia, 5-10 September, 1993;
to be published in the Proceedings.}
        {\rm and} M. BERGMANN }
\address{
         Institut f\"ur Theoretische Physik II  \\
         Ruhr-Universit\"at Bochum  \\
         D-44780 Bochum, Germany
   \\
         \\
         E-Mail:\ nicos@hadron.tp2.ruhr-uni-bochum.de
        }
\maketitle
\section{INTRODUCTION}
Insight into the nucleon substructure is being advanced remarkably by
studies~\cite{CZ84a,GS87a,KS87,COZ89a} of the quark momentum
distribution amplitudes, derived on the basis of perturbative Quantum
Chromodynamics (QCD)~\cite{LB80} supplemented by QCD sum
rules~\cite{CZ84b}. One of the most productive lines of research has
focused on the calculation of electroweak form factors of the nucleon
and its lowest
resonances~\cite{CZ84a,GS86,Car86,CP86,CGS87,JSL87,Ste89,CP88,Far88}.
A theoretical scheme which successfully describes such processes is
provided by the convolution formalism of Brodsky and Lepage~\cite{LB80}.
It assumes factorization of short- from large-distance effects to
separate out perturbative from nonperturbative contributions. All soft
gluon contributions of the form
$
 [\alpha_{\text{s}}\ln (Q^{2}/\mu^{2}) ]^{\gamma_ {\text{n} } }
$
are absorbed in the nucleon (or nucleon-resonance) distribution
amplitude
$
 \Phi_{\text{N}}(x_{\text{i}},Q^{2}).
$
This amplitude satisfies a renormalization-group equation which governs
the dynamical evolution of its $Q^{2}$-dependence. The solution can be
written as
\begin{equation}
  \Phi_{\text{N}}(x_{\text{i}},Q^{2})=
  \Phi_{\text{as}}(x_{\text{i}})\sum_{n=0}^{\infty}B_{\text{n}}
  \tilde \Phi_{\text{n}}(x_{\text{i}})\biggl(
  \frac{\alpha_{\text{s}}(Q^{2})}
  {\alpha_{\text{s}}(\mu^{2})}\biggr)^{\gamma_{\text{n}}},
\end{equation}
where
$\{\Phi_{\text{n}}\}_{n=0}^{\infty}$
are orthonormalized eigenfunctions of the interaction kernel within a
truncated basis of Appell polynomials of maximum degree $M$ and
$\Phi_{\text{as}}(x_{\text{i}})=120x_{1}x_{2}x_{3}$
is the asymptotic form of the nucleon distribution amplitude.
The corresponding (degenerate) eigenvalues $\gamma_{\text{n}}$ turn
out~\cite{Pes79} to be the anomalous dimensions of multiplicatively
renormalizable $I_{1/2}$ baryonic operators of twist three.
Because the $\gamma_{\text{n}}$ are positive fractional numbers
increasing with $n$, higher terms in this expansion are gradually
suppressed. Asymptotically, the $Q^{2}$ evolution of the distribution
amplitude is determined by the highest of the eigenvalues, i.e.,
$\gamma_{0}$. A basis including a total of $54$ eigenfunctions ($M=9$)
together with the associated normalization coefficients and anomalous
dimensions is given in \cite{BS93c}.

\section{CALCULATIONAL METHOD}

For the modeling of the nucleon distribution amplitude in terms of
its constituents it is unavoidable to employ nonperturbative methods,
such as QCD sum rules~\cite{CZ84b}, lattice gauge theory~\cite{RSS87} or
the direct diagonalization of the light-cone Hamiltonian within a
discretized light-cone setup~\cite{BP91}.

Inverting Eq.~(1), and making use of the orthonormalization of the
eigenfunctions $\Phi_{n}$, the (nonperturbative) expansion coefficients
$B_{\text{n}}$ at the renormalization point $\mu^{2}$ are given by
\begin{equation}
  B_{\text{n}}(\mu^{2})=\frac{N_{\text{n}}}
  {120}\int_{0}^{1}[dx]\tilde
  \Phi_{\text{n}}(x_{\text{i}})\Phi_{\text{N}}(x_{\text{i}}, \mu^{2})
\end{equation}
[$
  \int_{0}^{1} [dx] \equiv \int_{0}^{1} dx_{1} \int_{0}^{1-x_{1}}
  dx_{2} \int_{0}^{1} dx_{3} \delta (1-x_{1}-x_{2}-x_{3})
$]
and the ``renormalization-group improved'' coefficients
$
 B_{\text{n}}(Q^{2})
$
are then
\begin{equation}
  B_{\text{n}}(Q^{2}) = B_{\text{n}}(\mu^{2})
  \text{exp}
  \Biggl\{ -\int_ { \alpha_ {\text{s}} (\mu^{2})
                  } ^
                  { \alpha_ {\text{s}} (Q^{2})
                  }
  \frac {d\alpha} {\beta(\alpha )}
  \gamma_{\text{n}}(\alpha )\Biggr\}
  \approx
  B_ {\text{n}} (\mu^{2})
     \Biggl \{
             \frac {\ln (Q^{2}/\Lambda_{\text{QCD}}^{2})
             }
             {\ln (\mu^{2}/\Lambda_{\text{QCD}}^{2})
             }
     \Biggr \} ^{-\gamma_{\text{n}}}.
\end{equation}

A natural basis of the space of eigenfunctions of the evolution equation
is provided by Appell polynomials since they correspond to operators
with definite anomalous dimensions. Choosing as independent variables
$x_{1}$ and $x_{3}$, the moments of the nucleon distribution amplitude are
\begin{equation}
  \Phi_{\text{N}}^{(\text{i}0\text{j})}(\mu^{2}) =
  \int_{0}^{1}[dx]x_{1}^{\text{i}}x_{2}^{0}x_{3}^{\text{j}}
  \Phi_{\text{N}}(x_{i},\mu^{2}),
\end{equation}
so that Eq.~(2) becomes
\begin{equation}
  \frac{B_ {\text{n}} (\mu^{2})}{\sqrt{N_{\text{n}}}} =
  \frac{\sqrt{N_{\text{n}}}}{120}
  \sum_{\text{i,j}=0}^{\infty}a_{\text{ij}}^{\text{n}}\
  \Phi_{\text{N}}^{(\text{i}0\text{j})}(\mu^{2}),
\end{equation}
where the projection coefficients
$
 a_{\text{ij}}^{\text{n}}
$
are calculable to any order $M$~\cite{BS93c}.

To determine the moments
$
 \Phi_{\text{N}}^{(n_{1}n_{2}n_{3})},
$
a short-distance operator product expansion is performed at some
spacelike momentum $\mu^{2}$ where quark-hadron duality is
valid~\cite{CZ84b}. One considers
($z$ is a lightlike auxiliary vector with $z^{2}=0$)
\begin{eqnarray}
  \Biggl(iz\cdot \frac{\partial}{\partial z_{\text{i}}}
  \Biggr)^{n_{\text{i}}}
  \Phi_{\text{N}}(z_{\text{i}}\cdot p)\Bigg\vert_{z_{\text{i}=0}}
         & = &
   \prod_{\text{i}=1}^{3}
  \Biggl(iz\cdot \frac{\partial}{\partial z_{\text{i}}}
  \Biggr)^{n_{\text{i}}}
  \int_{0}^{1}[dx]\ e^ { -i\sum_{\text{i}=1}^{3}(z_{\text{i}}\cdot
  p)x_{\text{i}} }
  \Phi_{\text{N}}(x_{\text{i}}) \Bigg\vert_{z_{\text{i}=0}}
  \nonumber \\
         & = &
   (z\cdot p)^{n_{1}+n_{2}+n_{3}}
  \Phi_{\text{N}}^{(n_{1}n_{2}n_{3})}
\end{eqnarray}
and evaluates correlators of the form~\cite{CZ84a,KS87,COZ89a}
\begin{eqnarray}
  I^{\,(n_{1}n_{2}n_{3},m)}(q,z) & = & i\int_{}^{}
  d^{4}x \, e^{iq\cdot x}
  <\Omega\vert T\bigl (O_{\gamma}^{\,(n_{1}n_{2}n_{3})}(0)
  \hat O_{\gamma\prime}^{\,(m)}(x)\bigr )\vert\Omega>(z\cdot \gamma)_
  {\gamma \gamma\prime} \nonumber\\
  & & \\
  &  = &
  (z\cdot q)^{\,{n_{1}+n_{2}+n_{3}+m+3}}
  I^{\,(n_{1}n_{2}n_{3},m)}(q^{2})\;,
\nonumber
\end{eqnarray}
where the factor $(z\cdot \gamma )_{\gamma\gamma\prime}$ serves to
project out the leading-twist structure in the correlator, and
\begin{equation}
  O^{(n_{1}n_{2}n_{3})} =
  (z\cdot p)^{-(n_{1}+n_{2}+n_{3})}
  \prod_{i=1}^{3}(iz\cdot {\partial \over {\partial
  z_{\text{i}}}})^{n_{\text{i}}}
  O(z_{\text{i}}\cdot p)\big\vert _{z_{\text{i}}=0}
\end{equation}
are appropriate three-quark operators containing derivatives.
[Note that covariant derivatives acting on color-singlet operators equal
ordinary derivatives.] Their matrix elements
\begin{equation}
  <\Omega\vert O_{\gamma}^{(n_{1}n_{2}n_{3})}(0)\vert P(p)> =
  f_{N}(z\cdot p)^{\,{n_{1}+n_{2}+n_{3}+1}}N_{\gamma}\,
  O^{\,(n_{1}n_{2}n_{3})}
\end{equation}
are related to moments of the covariant distribution
amplitudes~\cite{HKM75} $V$, $A$, and $T$:
$
 \Phi_{\text{N}}(x_{\text{i}})=V(x_{\text{i}})-A(x_{\text{i}}),
$
$
 \Phi_{\text{N}}(1,3,2)+\Phi_{\text{N}}(2,3,1)=2T(1,2,3)
$
with
V(1,2,3)=V(2,1,3), A(1,2,3)=-A(2,1,3), and T(1,2,3)=T(2,1,3).
Here $f_{\text{N}}$ denotes the ``proton decay constant'' to be
determined from QCD sum rules~\cite{CZ84b}.

\section{MODELING THE NUCLEON DISTRIBUTION AMPLITUDE}

Lacking a complete understanding of the nonperturbative regime of QCD,
we turn to QCD sum rules and attempt to (re-)construct the nucleon
distribution amplitude from its moments. However, the problem of
determining an unknown (even non-negative) distribution from a given
{\it finite} set of its moments has no unique solution (see for example
\cite{Ste89} and references cited therein). Thus, in practice,
one uses the existing sets of QCD sum-rule constraints on the moments to
determine the expansion coefficients $B_{\text{n}}$ (c.f. Eq.~2) of
corresponding order within a truncated basis of Appell polynomials.
Theoretical constraints on the first $18$ moments of the distribution
amplitudes $(V-A)$ and $T$  of the nucleon and the $\Delta^{+}$-isobar
have been obtained in \cite{CZ84a,KS87,COZ89a} and \cite{CP88,Far88},
respectively. This program~\cite{Ste89} is attractive since the input is
reasonably well-defined and one does not rely on additional {\it ad hoc}
higher-order parameters, unspecified by theoretical constraints, which
must be extracted from the data. Physically, this truncated
representation of the distribution amplitude can be thought of in analogy
to a holographic image which is not destroyed when cut into pieces but
becomes rather less sharp~\cite{Ste92}.

It turns out that such models (see in~\cite{JSL87,Ste89,Ste92}) predict
approximately the right size and $Q^{2}$-evolution of
$G_{\text{M}}^{\text{p}}$, depending on the value of the scale parameter
$\Lambda_{\text{QCD}}$. On the other hand, the predictions for
$G_{\text{M}}^{\text{n}}$
are significantly model dependent: COZ-type models~\cite{COZ89a} predict
$\vert G_{\text{M}}^{\text{n}}\vert /G_{\text{M}}^{\text{p}}\le 0.5$
in connection with a small value of the electromagnetic transition form
factor $G_{\text{M}}^{*}$ (see~\cite{CGS87})---in contradiction with the
available data (see Fig.~3 below). Alternatively, the possibility
of having
$\vert G_{\text{M}}^{\text{n}}\vert \ll G_{\text{M}}^{\text{p}}$,
as frequently discussed in phenomenological data analyses~\cite{KK77},
is realized by the GS model~\cite{GS87a,GS86,Ste89} which gives for this
ratio the value $0.097$, and a sizeable value of $G_{\text{M}}^{*}$.
Although this model gives good agreement with the latest high-$Q^{2}$
SLAC data~\cite{Arn86}, some of the amplitude moments cannot match the
sum-rule requirements~\cite{CZ84a} in the allowed saturation
range~\cite{Ste89}. In addition, the calculated decay width of the
charmonium level ${}^{3}S_{1}$ into $p\bar p$ is much too low compared
with its experimental value~\cite{COZ89b}. Hence, either way, it is not
possible to reconcile the theoretical QCD sum-rule constraints with the
data because none of the models discussed so far gives a quantitative
account of the experimental form-factor and charmonium-decay data.

Perhaps understandably, in view of such distinctive models, the
conventional view has been one of fragmentation. But recent
research~\cite{SB92a,SB92b,BS93a,BS93b} suggests that, rather, all of
these models represent differentiated aspects of a single whole. Indeed,
it turns out to be possible to amalgamate the best features of COZ-type
and GS-type nucleon distribution amplitudes into a hybrid-like amplitude,
we termed the ``heterotic'' solution (see Fig.~1). As it will be more
transparent in the course of this talk, this novel amplitude seems
to have a foot in each of the previous models, lifting the disparity
between theory and experiment. This duality is also reflected in its
geometrical characteristics: the associated amplitudes with definite
permutation symmetry $V$ and $T$ are GS-like and COZ-like,
respectively~\cite{BS93b}. \par

\begin{figure}
\vspace{11 true cm}
\caption{ Optimized versions of the COZ- and GS-amplitudes
        in comparison with the novel heterotic amplitude.}
\end{figure}

This is achieved by recourse to a ``hierarchical'' treatment of the sum
rules~\cite{BS93a,BS93b} that accounts for the higher stability of the
lower-level moments~\cite{Ste89} and does not overestimate the
significance of the still unverified constraints~\cite{COZ89a} for the
third-order moments. Thus for each moment
$m_{\text{k}}$ (k=1,\ldots ,18),
we define
$
  \chi^{2}_{\text{k}}=(\chi^{2}_{\text{k},(\text{a})}+
  \chi^{2}_{\text{k},(\text{b})})
  \,[1\, - \,\Theta (m_{\text{k}}-M_{\text{k}}^{\text{min}})\Theta
  (M_{\text{k}}^{\text{max}}-m_{\text{k}})]
$
with
$
  \chi^{2}_{\text{k},(\text{a})}=min ( \vert
  M_{\text{k}}^{\text{min}}-m_{\text{k}}\vert,
  \vert m_{\text{k}}-M_{\text{k}}^{\text{max}}\vert )N_{\text{k}}^{-1},
$
where
$
 N_{\text{k}}=\vert M_{\text{k}}^{\text{min}} \vert
$
or
$
 \vert M_{\text{k}}^{\text{max}} \vert
$,
whether $m_{\text{k}}$ lies on the left-or on the right-hand side of the
corresponding sum-rule interval
($\chi^{2}_{\text{tot}}=\sum_{\text{k}}^{}\chi^{2}_{\text{k}}$)
and
\begin{equation}
\chi^{2}_{\text{k},(\text{b})}=
  \left\{\begin{array}{lll}
         100,   & 1\leq k\leq 3\\
         10,    & 4\leq k\leq 9\\
         1,     & 10\leq k\leq 18.
\end{array}
\right.
\end{equation}
The underlying assumption, parametrized in Eq.~(10) in terms of
(arbitrary) penalty points, is that contributions of higher-order terms
are either negligible or of minor importance relative to the second-order
terms and that including such terms only refines the initial second-order
ansatz~\cite{Ste89,SB92a}. Then the model space is also truncated at
states with bilinear correlations of fractional momenta and the pattern
of solutions found in this order should dominate the (orthonormalized)
Appell polynomial series at every order of truncation. This is quite
analog to approximating the field theory by truncating the Fock-space
expansion (Tamm-Dancoff method).

On the basis of those ideas, depending on the degree of matching with the
corresponding sum rule, a given solution appears as a local minimum of
the underlying $\chi^{2}$ criterion: lowest in the area of COZ-like
solutions, i.e., for
$
 0.45 \leq R \equiv
 \vert G_{\text{M}}^{\text{n}}\vert / G_{\text{M}}^{\text{p}}
 \leq 0.49.
$
(see Table~I).

\begin{table}
\caption{Theoretical parameters defining the nucleon distribution
         amplitudes discussed in the text.}
\begin{tabular}{lrrrrrrrrc}
 Model       & $B_{1}$\ \ &  $B_{2}$\ \ &   $B_{3}$\ \  &    $B_{4}$\ \ &
$B_{5}$\ \ &
$\vartheta$[deg]\  &  R\ \ & $\chi^{2}$\ & Symbol \cr
\tableline
 $Het     $ & 3.4437 &  1.5710 &   4.5937  &   29.3125 &    -0.1250  &    -1.89
 &    .104 &  33.48  &{\Large $\bullet$} \cr
 $Het^\prime $& 4.3025&  1.5920 &   1.9675  &  -19.6580 &     3.3531  &
24.44  &    .448 &  30.63 &{\Large $\bullet$} \cr
 $COZ^{\text{opt}}$ & 3.5268 &  1.4000 &   2.8736  &   -4.5227 &     0.8002  &
   9.13  &    .465 &   4.49 &$\blacksquare$    \cr
 $COZ^{\text{up}}$ & 3.2185 &  1.4562 &   2.8300  &  -17.3400 &     0.4700  &
5.83  &    .4881& 21.29 &$+$ \cr
 $COZ      $ & 3.6750 &  1.4840 &   2.8980  &   -6.6150 &     1.0260  &
10.16  &    .474 &  24.64 &$\Box$   \cr
 $CZ       $ & 4.3050 &  1.9250 &   2.2470  &   -3.4650 &     0.0180  &
13.40  &    .487 & 250.07 &$\blacklozenge$ \cr
 $KS^{\text{low}}$ & 3.5818 &  1.4702 &   4.8831  &   31.9906 &     0.4313  &
 -0.93  &    .0675&  36.27  &$\circ$ \cr
 $KS/COZ^{\text{opt}}$  & 3.4242 &  1.3644 &   3.0844  &   -3.2656 &     1.2750
 &     9.47  &    .453 &   5.66 &$\circ$    \cr
 $KS^{\text{up}}$ & 3.5935 &  1.4184 &   2.7864  &  -13.3802 &   2.0594   &
13.82  &    .482 & 40.38 &$\circ$ \cr
 $KS       $ & 3.2550 &  1.2950 &   3.9690  &    0.9450 &     1.0260  &
2.47  &    .412 & 116.35 &$\Diamond$ \cr
 $GS^{\text{opt}} $ & 3.9501 &  1.5273 &  -4.8174  &    3.4435 &     8.7534  &
  80.87  &    .095 &  54.95 &$\blacktriangle$ \cr
 $GS^{\text{min}} $ & 3.9258 &  1.4598 &  -4.6816  &    1.1898 &     8.0123  &
  80.19  &    .035 &  54.11 &$\blacktriangledown$ \cr
 $GS       $ & 4.1045 &  2.0605 &  -4.7173  &    5.0202 &     9.3014  &
78.87  &    .097 & 270.82 &$\bigtriangleup$ \cr
\end{tabular}
\end{table}

A global pattern of nucleon distribution amplitudes conforming with the
existing sets of sum rules~\cite{COZ89a,KS87} has been determined
in~\cite{BS93b} by systematically scanning the parameter space spanned
by the Appell decomposition coefficients $B_{\text{n}}$. The upshot of
this analysis is a dynamical arrangement of such solutions across an
orbit in $(B_{4},R)$ plane---ranging from COZ-like amplitudes to the
novel heterotic solution---depending on the value of
$
 R\equiv \vert G_{\text{M}}^{\text{n}}\vert /G_{\text{M}}^{\text{p}}
$
(see Fig.~2 and Table~I). These solutions seem to undergo a complete
metamorphosis as they pass from one end of the orbit to the other. For a
full discussion and classification of proposed model distribution
amplitudes, see~\cite{BS93b}.

\section{NUCLEON FORM FACTORS}

Employing analytical expressions derived in~\cite{Ste89}, the magnetic
form factor of the proton, calculated with the heterotic amplitude, is
shown in Fig.~3 in comparison with some previous models and recent
measurements (adapted from~\cite{Sil93}). When subsequent reference is
made to the heterotic model, it is to be understood that the values
$\Lambda_{\text{QCD}}=180 MeV$~\cite{GS87b} and
$\vert f_{\text{N}} \vert = (5.0 \pm 0.3) \times 10^{-3}
GeV^{2}$~\cite{COZ89a} have been used.
The form-factor evolution with $Q^{2}$ is due to the leading-order
parametrization of the effective coupling constant
$\alpha_{\text{s}}(Q^{2})$, whereas the evolution of the coefficients
$B_{\text{n}}$ has been neglected. Always an average value
$\bar \alpha_{\text{s}}(Q^{2})$ has been used
(peak approximation)~\cite{Ste89,SB92a}.

\begin{figure}
\vspace{8.7 true cm}
\caption{Pattern of nucleon distribution amplitudes matching the
         COZ sum-rule constraints (crosses) and such to a combined
         set of KS/COZ sum rules.}
\end{figure}

The results for the neutron magnetic form factor are shown in Fig.~3.
One observes that the disagreement among the various models is
substantial, while on the experimental side the available
data~\cite{Roc92} do not extend to the high-momentum region to phase out
a particular model. Noting that the theoretical curves correspond to the
helicity-conserving part of $G_{\text{M}}^{\text{n}}$, one appreciates
that COZ-like models may be in serious disagreement with the data.

Fig.~3 shows that the transition form factor, calculated with
the heterotic amplitude and $\Delta$ amplitudes from recent
QCD-analyses~\cite{CP88,Far88,SB92b}, is positive with a
seizable magnitude, though, smaller compared to the GS one.
The best agreement with the data is provided by a distribution amplitude
(denoted again ``heterotic'') derived by complying with the
CP~\cite{CP88} and FZOZ~\cite{Far88} sum-rule requirements {\it
simultaneously}~\cite{SB92b}. This solution satisfies all FZOZ
constraints and affords the best possible fit to the CP ones.
In particular, including the effect of perturbative (i.e., {\it
logarithmic}) $Q^{2}$ evolution of the expansion coefficients
$B_{\text{n}}$, the joined use of the heterotic amplitudes for the
nucleon and the $\Delta^{+}$ yields a form factor behavior which
conforms with the observed decrease of available data within their quoted
errors. To resolve more specifically whether $G_{\text{M}}^{*}$ decreases
at higher $Q^{2}$ values by a power-law fall-off (which would mean
dominance of its helicity non-conserving component~\cite{Sto93}) or
logarithmically~\cite{SB92b}, further precise measurements are required.

\begin{figure}
\vspace{10.5 true cm}
\caption{The curves are theoretical calculations of $F_{1}^{\text{p}}$,
         $G_{\text{M}}^{\text{n}}$, and $G_{\text{M}}^{*}$
         compared with available data.}
\end{figure}
A similar good agreement with the data is found also for the axial form
factors. In the $Q^{2}$ region where the calculations can be trusted,
we find~\cite{SB92a,Ste92}
${g_{\text{A}}(Q^{2})}/{G_{\text{M}}^{\text{p}}(Q^{2})}\approx 1.19$,
which compares well with the (extrapolated) experimental value
${g_{\text{A}}(Q^{2})}/{G_{\text{M}}^{\text{p}}(Q^{2})}\approx 1.35$.
If a dipole form
$g_{\text{A}}(Q^{2})=g_{\text{A}}(0)/(1+Q^{2}/{M_{\text{A}}^{2})^{2}}$
($g_{\text{A}}(0)=1.254\pm 0.006$) is assumed for the
$Q^{2}$ dependence of $g_{\text{A}}(Q^{2})$, then, at $Q^{2}\approx 10
GeV^{2}/c^{2}$, the heterotic model yields $M_{\text{A}}\approx
1.035 GeV$. This result actually coincides with the world-average value
$\bar M_{\text{A}}=1.032\pm 0.036 GeV$.
Analogously, one finds for
$
 G_{\text{A}}^{(\text{s})}(Q^{2})=G_{\text{A}}^{(\text{s})}(0)/
 (1+Q^{2}/{M_{\text{AS}}^{2})^{2}},
$
$M_{AS}=1.27 GeV$
with
$
 G_{\text{A}}^{(\text{s})}(0)=0.38
$
from $SU(6)$.

\section{CHARMONIUM DECAYS}

The study of exclusive charmonium decays into $p\bar p$ is also an
important tool in exploring the physics of the nucleon distribution
amplitudes and the correctness of proposed models. In Table II we show
some decay amplitudes which exhibit strong model dependence. These
calculations are made difficult by singularities and an elaborated
integration routine has been used to treat them properly~\cite{Ber93}.

\begin{table}
\caption{Decay amplitudes $M_{\text{i}}$ of charmonium states into
         $p\bar p$ extracted from various models.}
\begin{tabular}{rrrrrrr}
$(J^{\text{PC}}):\;\;M_{\text{i}}$ & Asympt. & CZ & COZ & KS & GS & Het. \\
\tableline
${}^3S_{1}\;\;(1^{--}):\;\;M_{0}$ & 1517.4   & 7545.6   & 8758.8 & 11484.0 &
928.8 & 13726.8  \\
${}^3P_{1}\;\;(1^{++}):\;\;M_{1}$ & 20086.56 & 28310.4  & 53625.6 & 94723.2 &
26366.4 & 99849.6  \\
${}^3P_{2}\;\;(2^{++}):\;\;M_{2}$ & 43099.2  & 246052.8 & 298123.2 & 416937.6 &
232632.0 & 515491.2 \\
\end{tabular}
\end{table}

Let us consider first the
${}^{3}P_{\text{J}}$ states with $J=1,2$. The branching ratio for the
decay of the $\chi_{\text{c}1}$ state into $p\bar p$ is given by
$
  BR\Bigl({{{}^3P_{1}\to p\bar p}\over {{}^3P_{1}\to\text{
  all}}}\Bigr)
  \approx {{0.75}\over{\ln({\bar M}/{\Delta})}}
  {{16\pi^{2}}\over{729}}{\Big \vert {{f_{\text{N}}}\over{{\bar
  M}^{2}}}
  \Big \vert}^{4}{M_{1}^{2}},
$
where $\bar M\approx 2 m_{\text{c}} \approx 3 GeV$ and $\Delta =0.4 GeV$
(the last value from~\cite{BGK76}). The nonperturbative content is
due to $f_{\text{N}}$ and the decay amplitude $M_{1}$ which involves
$\Phi_{\text{N}}$. The prediction of the heterotic model~\cite{SB92a}
(c.f., Table~II) is
$
 BR(^3P_{1}\to p\bar p/^3P_{1}\to \text{all})=0.77\times 10^{-2}\%
$,
which is in excellent agreement with the recent high-precision
experimental value~\cite{Arm92}
$(0.78\pm 0.10\pm 0.11)\times 10^{-2}$
of the E760 Colaboration at Fermilab. Analogously, the branching ratio
for the $\chi_{\text{c}2}$ state is
$
  BR\Bigl({{{}^3P_{2}\to p\bar p}\over {{}^3P_{2}\to\text{all}}}\Bigr)
  \approx 0.85 (\pi\alpha_{\text{s}})^{4}
  {{16}\over{729}}{\Big \vert {{f_{\text{N}}}\over{{\bar M}^{2}}}
  \Big \vert}^{4}{M_{2}^{2}},
$
and setting $\alpha_{\text{s}}(m_{\text{c}})=0.210\pm 0.028$ (see third
paper of~\cite{BGK76}), we obtain
$
 BR({{}^3P_{2}\to p\bar p}/{{}^3P_{2}\to\text{all}})=
 0.89\times 10^{-2}\%
$
in excellent agreement with the high-precision E760-value~\cite{Arm92}
$(0.91\pm 0.08\pm 0.14)\times 10^{-2}\%$.

Similar considerations apply also to the charmonium decay of the level
${}^{3}S_{1}$.
The partial width of $J/\psi$ (or $\chi_{\text{c}0}$) into $p\bar p$ is
$
  \Gamma({}^3S_{1}\to p\bar p) =
  (\pi\alpha_{\text{s}})^{6}{{1280}\over{243\pi}}
  {{{\vert f_{\psi}\vert}^{2}}\over{\bar M}}
  {\Big \vert {{f_{\text{N}}}\over{{\bar M}^{2}}}
  \Big \vert}^{4}{M_{0}^{2}},
$
where $f_{\psi}$ determines the value of the ${}^3S_{1}$-state wave
function at the origin. Its value can be extracted from the leptonic
width $\Gamma({}^3S_{1}\to e^{+}e^{-}) = (5.36\pm 0.29)
keV$~\cite{HP92} via the Van Royen-Weisskopf formula. The result is
$\vert f_{\psi}\vert =409 MeV$ with
$m_{J/\psi}=3096.93 MeV$~\cite{PDG92}.
Then, using the previous parameters, it follows that
$\Gamma({}^3S_{1}\to p\bar p) = 0.14 keV$. From experiment~\cite{PDG92}
it is known that $\Gamma(p\bar p)/\Gamma_{\text{tot}}=
(2.16\pm 0.11)\times 10^{-3}$ with $\Gamma_{\text{tot}}=(68\pm 10) keV$,
so that
$\Gamma({}^3S_{1}\to p\bar p) = 0.15 keV$ in perfect agreement with
the model prediction.
The corresponding branching ratio is
$
  BR ( {}^{3}S_{1}\to p\bar p/ {}^{3}S_{1}\to\text{all} )=
  1.62\times 10^{-3},
$
with
$\Gamma_{\text{tot}}=(85.5{+6.1\atop -5.8}) keV$.
To effect the quality of these predictions, we quote the results
for the COZ amplitude calculated by the same authors~\cite{COZ89b}:
[Note that they use the rather arbitrary value $\alpha_{\text{s}}=0.3$.]
$
 BR(^3P_{1}\to p\bar p/^3P_{1}\to \text{all})=0.50\times 10^{-2}\%
$,
$
 BR({{}^3P_{2}\to p\bar p}/{{}^3P_{2}\to\text{all}})=
 1.6\times 10^{-2}\%
$, and
$\Gamma({}^3S_{1}\to p\bar p) = 0.34 keV$.
On the other hand, the GS model gives, e.g.,
$\Gamma({}^3S_{1}\to p\bar p) = 0.56\times 10^{-3} keV$
($\alpha_{\text{s}}=0.21$) and
$\Gamma({}^3S_{1}\to p\bar p) = 0.26\times 10^{-2} keV$
($\alpha_{\text{s}}=0.30$), in contradiction to the measured value.
However, one can force agreement between this model and experiment by
increasing $\alpha_{\text{s}}$ to the (arbitrary) value $0.39$.

\section{CONCLUDING REMARKS}

The picture that emerges from the studies discussed above---all in the
pursuit of a deeper understanding of the nucleon substructure---is
analytically incomplete but technically sufficient to establish
a good agreement between theoretical results and experimental data,
supporting the validity of the assumptions made in the model-dependent
analyses.
In particular, recent progress~\cite{LS92,JK93} in Sudakov-suppression
techniques provides support for the conjectured infrared protection of
the perturbative picture.
[We should mention, however, the objections raised in~\cite{IL84,Rad91}.]
In particular, approximations notwithstanding, the heterotic
model seems to be capable of reproducing the observed phenomena with
remarkable accuracy. While effects of higher-order eigenfunctions on the
nucleon distribution amplitude itself are claimed to be
large~\cite{Sch89,HEG92}, the agreement with the data is
actually not improved~\cite{BS93c}. Furthermore, at relatively large
distances probed in present experiments, still uncalculable contributions
of higher twists, i.e., multi-parton distribution amplitudes, are
presumably more significant than higher-order terms of the Appell
polynomial series which correspond to higher correlations of the
fractional momenta of the valence quarks. Exactly which effects dominate
is not well understood at present and deserves further investigations.
\acknowledgements
I wish to thank the organizers for the warm hospitality extended to me
during the meeting. This work was supported in part by the Deutsche
Forschungsgemeinschaft.

\end{document}